\documentclass[fleqn,usenatbib]{mnras}

\usepackage{newtxtext,newtxmath}

\usepackage[T1]{fontenc}

\DeclareRobustCommand{\VAN}[3]{#2}
\let\VANthebibliography\thebibliography
\def\thebibliography{\DeclareRobustCommand{\VAN}[3]{##3}\VANthebibliography}

\usepackage{graphicx}	
\usepackage{amsmath}	
\usepackage{comment}
\usepackage{ulem}
\usepackage{csvsimple}
\usepackage{tabularx}
\usepackage{geometry}

\usepackage{longtable}
\usepackage{booktabs}
\usepackage{multirow}
\usepackage{comment}

\usepackage{tikz,xcolor,hyperref}
\definecolor{lime}{HTML}{A6CE39}
\DeclareRobustCommand{\orcidicon}{%
	\begin{tikzpicture}
	\draw[lime, fill=lime] (0,0) 
	circle [radius=0.16] 
	node[white] {{\fontfamily{qag}\selectfont \tiny ID}};
	\draw[white, fill=white] (-0.0625,0.095) 
	circle [radius=0.007];
	\end{tikzpicture}
	\hspace{-2mm}
}

\foreach \x in {A, ..., Z}{%
	\expandafter\xdef\csname orcid\x\endcsname{\noexpand\href{https://orcid.org/\csname orcidauthor\x\endcsname}{\noexpand\orcidicon}}
}

\title[QPO detection in PKS 0521-36]{Detection of gamma-ray quasi-periodic oscillations in non-blazar AGN PKS 0521-36}

\author[A. Sharma et al.]{
Ajay Sharma\orcidA,$^{1}$\thanks{ajjjkhoj@gmail.com }
Raj Prince\orcidB,$^{2}$\thanks{raj@cft.edu.pl}
Debanjan Bose\orcidC$^{3}$\thanks{debaice@gmail.com}
\\
$^{1}$ S. N. Bose National Centre for Basic Sciences, Block JD, Sector III, Salt Lake, Kolkata 700 106, India\\
$^{2}$ Center for Theoretical Physics, Polish Academy of Sciences, Al. Lotnikow, 32/46, Warsaw, Poland\\
$^{3}$ School of Astrophysics, Presidency University, Kolkata 700073, India
}

\date{Accepted XXX. Received YYY; in original form ZZZ}

\pubyear{2015}

\begin{document}
\label{firstpage}
\pagerange{\pageref{firstpage}--\pageref{lastpage}}
\maketitle

\begin{abstract}
Quasi-periodic oscillations (QPOs) have been detected in many Fermi-detected bright blazars. In this letter, we report multiple QPOs detected in a non-blazar AGN PKS 0521-36 searched over the entire 15 years of Fermi-LAT data. QPOs are detected at 268 days, at 295 days, and at 806 days timescales with more than 3$\sigma$ significance. The QPO detected at 806 days happens to be the third harmonic of QPO at 268 days. The time scales are consistent in both Lomb-Scargle and Wavelet analysis. Furthermore, the Gaussian Process modeling of the light curve is performed with stochastically driven damped harmonic oscillator (SHO) and damped random walk (DRW) modeling to uncover the presence of QPOs. The constructed power spectral density (PSD) exhibits two QPOs, with observed timescales of approximately 283 days and 886 days. This is the first non-blazar AGN where the long-term QPO is detected. Earlier studies show this source has a weak beamed jet. The exact cause for these QPOs remains unclear. We also assembled the $\gamma$-ray QPO detected in various blazar and tested the QPO time scale dependent on the black hole mass. No significant correlation is found.
\end{abstract}




\section{Introduction}

It is widely acknowledged that a small fraction of galaxies harbor Active Galactic Nuclei (AGN) at their centers, which are powerful emitters, and serve as potent sources of electromagnetic (EM) radiation. AGN are believed to derive their power from the process of accretion onto supermassive black holes and some of the AGN host relativistic jets perpendicular to the disk plane. They are the source of variable emissions that span from radio to very high-energy gamma-rays even higher. The high-energy component of this radiation is generally attributed to its origin within the powerful relativistic jets covering the MeV to TeV energy range.\par
The \textit{Fermi}-LAT telescope has achieved successful detection of gamma-ray from a class of AGNs, specifically those of the blazar type. The blazars are characterized by the presence of a relativistic jet oriented at a narrow angle ($ \theta < 5^{\circ}$) relative to the observer's line of sight \citep{2015ApJ...810...14A}. As a consequence of this alignment, the observed emission is strongly Doppler-boosted. The prevailing consensus is that the variability observed in AGNs, particularly in blazars, is inherently stochastic. This means that a power-law model across a broad range of temporal frequencies can effectively characterize the power spectral density (PSD). Blazars are known for their remarkable flux variability, which occurs rapidly and spans an extensive range of timescales, from mere minutes to several years. The continuum emissions from AGNs across various energy bands offer to search for multifrequency variability. Such studies provide valuable insights into the locations and sizes of these emissions and the underlying physical processes responsible for emissions. Conducting timing analysis, such as the exploration of periodic oscillations, would provide the necessary information on the intrinsic characteristics of these sources. The quest for QPOs with the Fermi-LAT within the gamma-ray domain of blazars has been an active area of research because it covers more than 15 years of baseline and provides a uniform sampling of the light curve. Based on observations made by the Fermi-LAT instrument, there have been reports suggesting the presence of possible QPOs of various time scale in more than 20 blazars, e.g. \citep{2014ApJ...793L...1S, 2015ApJ...810...14A, 2016AJ....151...54S, 2017MNRAS.471.3036P, sandrinelli2017gamma, zhang2017revisiting, zhang2017gamma, zhang2017possible, zhou201834, bhatta2019blazar, penil2020systematic, zhang2020searching, banerjee2023detection,  Das_2023, 2023arXiv230811317P}. However, the origins of $\gamma$-ray QPOs continue to be a subject of debate within the scientific community. It's worth noting that a considerable portion of these detections only spans a modest 2-4 cycles, potentially leading to overestimation in their significance.\par
The fourth catalog of AGNs detected by Fermi-LAT includes numerous non-blazar AGNs \citep{ajello2020fourth}, many of these sources, spanning different AGN classes, are variable. Detailed investigations have been conducted on certain sources belonging to the NLSy1s category, a distinct class of AGNs \citep{d2013multifrequency, d2015most, d2016awakening, d2019relativistic, paliya2019general}. NGC 1275, a radio galaxy (RG), has been very well observed by Fermi-LAT, resulting in abundant data and detailed investigations \citep{baghmanyan2017rapid, sahakyan2018fermi, tanada2018origins, Tomar_2021}. Unlike the typically bright blazars, most non-blazar AGNs do not exhibit sufficiently robust sampling in their long-term $\gamma$-ray light curves because of the low significant detection of gamma-ray in certain periods. However, few exceptions, such as NGC 1275 and PKS 0521-36, offer unique opportunities for comprehensive investigations into their $\gamma$-ray properties. A detailed study of these sources, focusing on their $\gamma$-ray properties, has the potential to yield valuable insights into emission mechanisms. Furthermore, by combining the study of non-blazar AGNs with that of blazars, we can enhance our understanding of the intricate physics governing AGN jets.

PKS 0521-36 is a non-blazar AGN detected by Fermi-LAT in gamma-ray along with many broad emission lines in optical and UV as reported in earlier studies \citep{1986ApJ...302..296K, 1995A&A...303..730S, 2017MNRAS.470L.107J}. The radio morphology of the source has been explored in detail in the past where the optical/near IR/sub-mm jet was found to be well aligned with the radio jet \citep{Scarpa_1999, 2009A&A...501..907F, 2016A&A...586A..70L}. In an earlier study by \citet{1995A&A...303..730S} the PKS 0521-36 was classified as flat spectrum radio quasars (FSRQ). The recent study done by \citet{2017MNRAS.470L.107J} has discovered an S-shape structure possibly caused by the interaction between the jet and the ISM at the kpc scale. A QPO search is done by \citet{Zhang_2021} using a part of the gamma-ray light curve from MJD 56317 -- 58447 and they detected a strong QPO signature at 400 days. However, In our study, we searched for the QPO feature in a 15-year-long light curve and three significant QPO peaks were observed at different time scales.
\par

In this work, we explore the $\gamma$-ray lightcurve observed by LAT since August 2008 and observed three quasi-periodic oscillations at periods of $\sim$268 days, $\sim$295 days, and $\sim$806 days by employing advanced mathematical techniques like Lomb-Scargle periodogram (LSP) and weighted-wavelet (WWZ) technique. In addition, we incorporated the Gaussian Process modeling such as stochastically driven damped harmonic oscillator (SHO) and damped random walk (DRW) in our investigation and discovered two QPO features with a period of $\sim$283 days and $\sim$886 days. In section 2, we provide the details about the gamma-ray analysis followed by the methodology used for QPO tests in sections 3 and 4. In section 5, we discuss our results and finally present the conclusion in section 6.

\section{OBSERVATIONS AND DATA REDUCTION}
We collected the Fermi-LAT \citep{atwood2009large} data spanning a 15-year timeframe, between MJD 54683 to MJD 60187 (Aug 2008 - 2023).
The observations were conducted within the energy range of 0.1 to 300 GeV. The analysis of the data follows the standard criteria for point-source analysis. Utilizing the Fermi Science tool package\footnote{\url{http://fermi.gsfc.nasa.gov/ssc/data/analysis/documentation/}} provided by the Fermi Science Support Center, we employed the Pass 8 LAT database. The region of interest (ROI) chosen for our analysis encompasses a circular region of 10$^{\circ}$ radius centered at PKS 05231-36. 
The data were filtered by the task \textsc{gtselect} with "evclass=128, evtype=3".\par

The data with zenith angle $> 90^{\circ}$ were filtered out to avoid any contamination originating from the Earth's Limb. To enhance the data quality, we applied a standard filtering procedure using the \textsc{gtmktime} tool to isolate the high-quality data within good time intervals (GTIs). This filtering was achieved by applying the standard criteria with recommended filter expression \texttt{$\text{(DATA\_QUAL > 0) \&\& (LAT\_CONFIG == 1)}$}. Using the \textsc{gtltcube} and \textsc{gtexposure} tools, we calculated the integrated livetime as a function of sky position and off-axis angle and exposure, respectively. The galactic and extra-galactic diffuse background emissions were modeled using two files: $\text{gll\_iem\_v07.fits}$\footnote{\url{https://fermi.gsfc.nasa.gov/ssc/data/access/lat/BackgroundModels.html}} and $\text{iso\_P8R3\_SOURCE\_V3\_v1.txt}$\footnote{\url{https://fermi.gsfc.nasa.gov/ssc/data/access/lat/BackgroundModels.html}}. The instrumental response function "$\text{P8R3\_SOURCE\_V3}$" was adopted in data processing. The unbinned likelihood analysis was performed, as described in the official unbinned likelihood tutorial\footnote{\url{https://fermi.gsfc.nasa.gov/ssc/data/analysis/scitools/likelihood_tutorial.html}}, using the \textsc{gtlike} tool \citep{cash1979parameter, mattox1996likelihood} that provided the significance of each source within the ROI including the source of interest in the form of test statistics (TS $\sim \sigma^2$; \citealt{mattox1996likelihood}). We then removed the sources with low TS i.e. below TS = 9.
We generated a 10-day binned lightcurve by integrating the source fluxes for the bins where $\text{TS} > 9 (\sim 3 \sigma)$, in the energy range 0.1 - 300GeV. The generation process involved an unbinned likelihood analysis method. Notably, parameters for sources positioned beyond 10$^{\circ}$ from the center of the ROI were fixed, while those $\le$10$^{\circ}$ range were left unrestricted and allowed to vary freely. To produce the lightcurve, we employed the \textsc{enrico} software\footnote{\url{https://enrico.readthedocs.io/en/latest/}} developed by \citep{sanchez2013enrico} and used the LAT source catalog $\text{gll\_psc\_v30.fit}$\footnote{\url{https://fermi.gsfc.nasa.gov/ssc/data/access/lat/12yr_catalog/}}.\par

The gamma-ray spectral energy distribution (SED) is also produced and fitted with a spectral model of the form of the LogParabola (appendix Figure \ref{Fig: GAMMA-RAY SED}) which is given as, 
\begin{equation}
    \frac{dN}{dE} = N_0 \left( \frac{E}{E_0}\right)^{-[\alpha + \beta log(E/E_b)]}
\end{equation}
, where $\alpha$ and $\beta$ represent the spectral index and curvature parameter, respectively, at the break energy ($E_b$). The normalization factor is denoted by $N_0$. The best-fitting spectral parameters values are, the $\alpha$ = $2.38\pm 0.014$, $\beta = 0.069\pm0.008$, and $N_0 = 0.5\pm0.006$. An integrated flux from the fitting is $(1.04\pm0.017)\times 10^{-7}$ ph cm$^{-2}$ s$^{-1}$. The Test Statistic is defined as TS = 2$\Delta$log(likelihood) = 2log($\frac{L}{L_0}$), where $L$ and $L_0$ are the maximum likelihood of the model with a point source at the target location and maximum likelihood value fitted by the background model, respectively. The fitting yields a total test statistic value of 15091.88.

\section{Methodologies}

We adopted several quantitative tests to identify Quasi-Periodic Oscillations (QPOs) in the gamma-ray light curve of the source. Specifically, we utilized the "Lomb-Scargle Periodogram (LSP)" and the "weighted wavelet Z-transform (WWZ)" analysis techniques. In the subsequent subsection, we will elaborate on our findings and the outcomes of applying these methods.

\begin{figure*}
    \centering
    \includegraphics[width=0.7\textwidth]{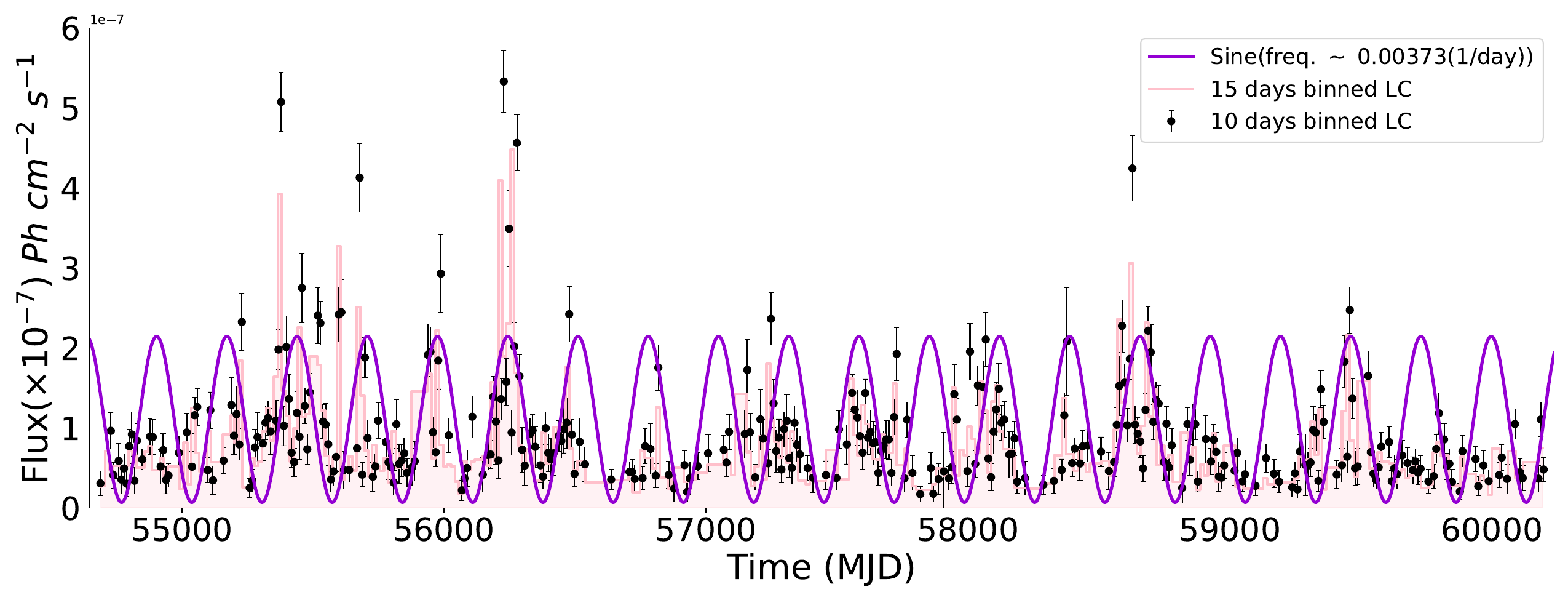}
    \caption{The Fermi-LAT light curve of the AGN PKS 0521-36 over the entire duration ($\sim$15yr). A 10-day binned lightcurve is depicted by black points and a 15-day binned lightcurve is illustrated using a step-bar in pink, all the data points are with TS $>$ 9. A sinusoidal function (purple solid line) characterized by an oscillation frequency of $\sim$0.00373 $\mathrm{d^{-1}}$ (corresponds to a period of 268 days), is plotted on top of the light curves to illustrate the possible periodicity in the light curve.}
    \label{Fig-1}    
\end{figure*}

\begin{figure*}
    \centering
    \includegraphics[width=0.8\textwidth]{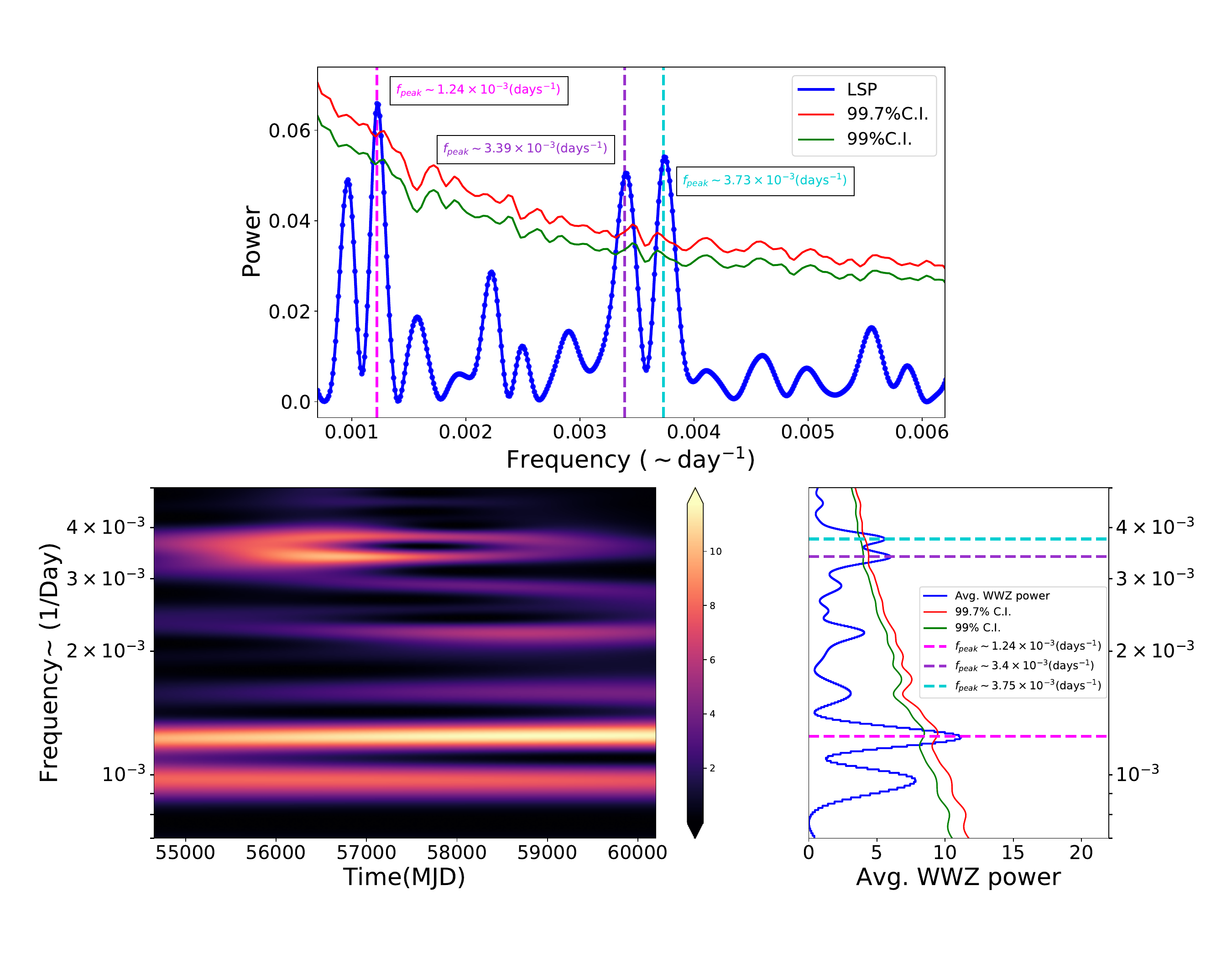}
    \caption{The QPO tests with LSP and WWZ are shown here. Top Panel: The LSP is depicted by the solid blue line, corresponding to the $\gamma$-ray lightcurve, with 99.7$\%$ and 99$\%$ significance lines in red and green color, respectively. Notably, the LSP exhibits peaks at $\sim$0.00124 $\text{d}^{-1}$, $\sim$0.00339 $\text{d}^{-1}$, and $\sim$0.00373 $\text{d}^{-1}$, with a significance levels surpassing 3$\sigma$. Bottom Left: The WWZ map of the $\gamma$-ray light curve is displayed, where the color bar indicates varying wavelet power levels. Clear concentrations of power around the frequencies of $\sim$0.00124 $\text{d}^{-1}$ (806 days), $\sim$0.0034 $\text{d}^{-1}$ (294 days), and $\sim$0.00375 $\text{d}^{-1}$ (266 days) strongly suggest the potential existence of periodicities. Bottom Right: The solid blue line represents the average wavelet power across time, while the red and green solid lines denote the 99.7$\%$ and 99$\%$ level of significance, respectively. This method provides additional confirmation of the statistical significance of the observed QPO frequencies. 
    }
    \label{fig-2}
\end{figure*}

\subsection{Lomb-Scargle periodogram}
The Lomb-Scargle periodogram method stands out as one of the most commonly employed techniques for detecting potential periodic patterns within time series data \citep{lomb1976least, scargle1982studies}. It represents a variant of the traditional discrete Fourier transform (DFT), with a notable advantage: it excels when dealing with unevenly sampled data. This method possesses the remarkable ability to reduce the impact of irregularities in data collection by iteratively fitting sinusoidal waves to the data. This fitting process effectively reduces the impact of the noise component and provides an accurate measurement of the detected periodicity. In this study, we computed the LSP using the \textsc{lomb-scargle}\footnote{\url{https://docs.astropy.org/en/stable/timeseries/lombscargle.html}} class provided by \textsc{astropy}.\par
In addition, it is conventional to test the presence of periodicity using the \textsc{generalized lomb-scargle periodogram}\footnote{\url{https://pyastronomy.readthedocs.io/en/latest/pyTimingDoc/pyPeriodDoc/gls.html}} (GLSP). This approach takes into consideration the measurement uncertainties during the analysis. The outcome of this analysis provides us with evidence of periodicity at the same periods, lending further support to our findings. The result of LSP is shown in Figure~\ref{fig-2}.

\subsection{Weighted wavelet Z-transform}

In the quest to uncover potential periodic patterns in blazar lightcurves, another widely adopted approach is the WWZ method. While the standard LSP technique is valuable, it has its limitations. LSP attempts to fit a sinusoidal profile over the entire span of observation, not taking into account the fact that the observed features coming from real astrophysical events may be time-dependent. This is where WWZ steps in as a complementary tool, offering a more comprehensive view of periodicity in the time series data. The Wavelet transform method has proven to be a more suitable tool for such studies. It involves convolving the light curves with the time and frequency-dependent kernel and attempts to localize the periodicity feature in time and frequency space. In the context of studying the evolution of QPO features over time, this method emerges as a powerful tool, allowing us to delve into the intricate details of how these oscillations gradually develop, evolve across the frequency space, and eventually fade away over time \citep{foster1996wavelets}.\par
In wavelet analysis, we used the abbreviated Morlet kernel \citep{grossmann1984decomposition} which has the following functional form:

\begin{equation}
    f[\omega (t - \tau)] = \exp[I \omega (t - \tau) - c \omega^2 (t - \tau)^2]
\end{equation}
and the corresponding WWZ map is given by,
\begin{equation}
    W[\omega, \tau: x(t)] = \omega^{1/2} \int x(t)f^* [\omega(t - \tau)] dt
\end{equation}
Here, $f^*$ is the complex conjugate of the wavelet kernel f, $\omega$ is the frequency, and $\tau$ is the time-shift. This kernel acts as a windowed DFT, where the size of the window is determined by both the parameters $\omega$ and a constant \textit{c}. The resulting WWZ map offers a notable advantage, it not only identifies dominant periodicities but also provides insights into their duration over time.\par
For this study, we use publicly available Python code\footnote{\url{https://github.com/eaydin/WWZ}} to generate the WWZ map. The obtained WWZ map shows a decipherable concentration of power around 0.00124 $\mathrm{d^{-1}}$ ($\sim$806 days). We also detect a significant concentration of power at 0.0034 $\mathrm{d^{-1}}$ ($\sim$295 days) and 0.0037 $\mathrm{d^{-1}}$ ($\sim$268 days) within the time domain of MJD 55000-59000 (Figure~\ref{fig-2}).

\subsection{Significance estimation}

The characteristic red-noise type variability pattern in temporal frequency serves as a distinctive hallmark of AGN or Blazars. The periodogram is usually represented as a power spectral density (PSD) of a form $P(\nu) \sim A \nu^{-\beta}$, where the temporal frequency is represented by $\nu$ and $\beta > 0$ represents the spectral slope. To gauge the statistical significance of the periodic feature, we employed the approach developed by \cite{emmanoulopoulos2013generating}. The methodology involved modeling the observed PSD using a power-law model. Subsequently, we conducted Monte Carlo simulation wherein we generated 1000 light curves using \textit{$\mathrm{DELightcurveSimulation}$}\footnote{\url{https://github.com/samconnolly/DELightcurveSimulation}} code, each mimicking the characteristics of the original data, including the best-fitting power-law model slope ($\beta$ = 0.48$\pm$0.16) and similar properties in terms of flux distribution.\par
As mentioned in the preceding subsection, we detected a notable peak and two transient peaks in both the LSP and the average WWZ. From LSP analysis, the periodicity feature at 0.00124 $\mathrm{d^{-1}}$ ($\sim$806 days)  exhibited a significance level exceeding 99.7$\%$. Similarly, the other two observed periodic features at 0.00339 $\mathrm{d^{-1}}$ ($\sim$295 days) and 0.0037 $\mathrm{d^{-1}}$ ($\sim$268 days) also displayed significance levels surpassing 99.7$\%$. The significance level of the observed peaks of WWZ power is estimated in the same way, periodicities detected at 0.00124 $\mathrm{d^{-1}}$ ($\sim$806 days), 0.0034 $\mathrm{d^{-1}}$ ($\sim$295 days), and 0.00375 $\mathrm{d^{-1}}$ ($\sim$266 days) are found to have significance level $ > 99.7\% $. The results are shown in Figure~\ref{fig-2}. 

\section{Stochastic process}
In addition to approaches such as LSP and WWZ for analyzing astronomical variability, an alternative methodology involves employing Gaussian process modeling of light curves in the temporal domain. One notable Gaussian process model is the continuous-time autoregressive moving average model developed by \cite{kelly2014flexible}, which has been applied to $\gamma$-ray variability of blazars \citep{kelly2009variations, kelly2014flexible, li2018new, goyal2018stochastic, ryan2019characteristic, tarnopolski2020comprehensive, yang2021gaussian, zhang2021quasi, zhang2022characterizing, sharma2024probing}. In our analysis, we adopted a newly developed Gaussian process model implemented in the \textsc{celerite} package \citep{foreman2017fast}. This method requires users to define a specific and stationary kernel function.

\subsection{DRW Model}
CARMA(p, q) processes \citep{kelly2014flexible} are defined as the solutions to the following stochastic differential equation:

\begin{equation}
\begin{split}
\frac{d^p y(t)}{dt^p} + \alpha_{p-1}\frac{d^{p-1}y(t)}{dt^{p-1}}+...+\alpha_0 y(t) =\\
\beta_q \frac{d^q \epsilon(t)}{dt^q}+\beta_{q-1}\frac{d^{q-1}\epsilon(t)}{dt^{q-1}}+...+\beta_0 \epsilon(t),
\end{split}
\end{equation}

where, y(t) is a time series, $\epsilon$(t) is a continuous time white-noise process, $\alpha^*$ and $\beta^*$ are the coefficients of AR and MA models, 
respectively. Parameters p and q determine the order of AR (Auto-regression) and MA (Moving-average) models, respectively. In the special case where the model's parameters are set to p=1 and q=0, representing a CAR(1) (continuous auto-regressive) model, often referred to as the Ornstein–Uhlenbeck process. In astronomical literature, the CAR(1) model is widely recognized as a DRW (Damped Random Walk) process and can be described by the stochastic differential
equation of the following form \citep{kelly2009variations, moreno2019stochastic},

\begin{equation}
    \left[ \frac{d}{dt} + \frac{1}{\tau_{DRW}} \right]y(t) = \sigma_{DRW} \epsilon(t)
\end{equation}

where $\tau_{DRW}$ is the characteristic damping time-scale of DRW process and
$\sigma_{DRW}$ is denoting the amplitude of random perturbations. To model the DRW process, the covariance function for this process is written as
\begin{equation}
    k(t_{nm}) = a.exp(-t_{nm}c),
\end{equation}
where $t_{nm} = | t_n -t_m|$ is the time lag, and $a = 2 \sigma_{DRW}^2$ and $c = \frac{1}{\tau_{DRW}}$. The PSD is written as:
\begin{equation}
    S(\omega) = \sqrt{\frac{2}{\pi}} \frac{a}{c} \frac{1}{1 + (\frac{\omega}{c})^2}
\end{equation}

The PSD of DRW is a broken PL form, where the broken frequency $f_b$ corresponds to the damping timescale $\tau_{DRW} = \frac{1}{2\pi f_b}$.

\subsection{SHO Model}
In our investigation, a stochastically driven damped simple harmonic oscillator (SHO) is used to model the light curve. The differential equation for this system is described as follows \citep{foreman2017fast} :

\begin{equation}
    \left[ \frac{d^2}{dt^2} + \frac{\omega_0}{Q}\frac{d}{dt} + \omega_{0}^{2} \right]y(t) = \epsilon(t)
\end{equation}

where $\omega_0$ is the frequency of the undamped oscillator, Q is the quality factor of the oscillator, and $\epsilon$(t) is a stochastic driving force. If the $\epsilon$(t) is white noise, the PSD of this process is given as

\begin{equation}
    S(\omega) = \sqrt{\frac{2}{\pi}} \frac{S_0 \omega_{0}^4}{\left( \omega^2 - \omega_{0}^2\right)^2 + \frac{\omega^2 \omega_{0}^2}{Q^2}}
\end{equation}
where $S_0$ is proportional to the power at $\omega = \omega_0$, $S(\omega_0) = \sqrt{\frac{2}{\pi}}S_0 Q^2$.

\subsection{Model Selection}

The Bayesian information criterion (BIC) serves as a means of capturing the information lost when employing models to characterize the underlying data-generation processes. Additionally, it provides insight into the comparative quality of different models applied to a dataset. The efficacy of a model in describing underlying processes is considered higher when less information is lost. To perform the model selection, we compute the BIC 

\begin{equation}
    BIC = -2 \textit{$\mathcal{L}^*$} + K\ logN
\end{equation}

where $\mathcal{L}^*$ is the value of maximum likelihood, K is the number of parameters, and N is the number of data points. A preferred model should have minimum BIC.\par
We characterized the $\gamma$-ray variability by employing various combinations of SHO and DRW terms. The kernels corresponding to these model combinations are as follows:

\begin{align}
    k(\tau) &= k(\tau ; S_0, Q, \omega_0) \times n \\
    k(\tau) &= k(\tau ; S_0, Q, \omega_0) + n \\
    k(\tau) &= k(\tau ; a, c) + k(\tau ; S_0, Q, \omega_0) \times n 
\end{align}

where n represents distinct SHO terms, constrained to be $\le$ 4. We selected these specific functional forms for kernels to identify the optimal model for the light curve through a comparison based on the BIC. We conducted the fitting process using \textsc{celerite2}\footnote{\url{https://celerite2.readthedocs.io/en/latest/}} and implemented the Markov chain Monte Carlo (MCMC) algorithm provided by the \textsc{emcee}\footnote{\url{https://emcee.readthedocs.io/en/stable/}} package. We determined the Maximum a posteriori (MAP) parameters using the nonlinear optimizer \textsc{L-BFGS-B}\footnote{\url{https://docs.scipy.org/doc/scipy/reference/optimize.minimize-lbfgsb.html}}, implemented by the \textsc{Scipy} project. We computed the maximum likelihood for each optimization iteration through 100 runs, minimizing effects arising from the instability of the \textsc{L-BFGS-B} algorithm. The maximum value obtained among these runs was utilized to calculate the BIC. We executed this procedure for each model, and the best model was identified based on the minimum BIC. In the celerite modeling, we ran the emcee sampler with 32 parallel walkers for a burn-in phase of 5000 steps, followed by 10000 steps as the product of MCMC.

\section{Result and Discussion}
PKS 0521-36 is an active galactic nucleus but quite bright and variable in $\gamma$-ray and a broadband study done in 2015 by \cite{10.1093/mnras/stv909} suggests that the viewing angle of the jet is between 6 to 15 deg. \cite{Zhang_2021} have detected a 400 days quasi-periodic oscillation in 5.8 years of Fermi-LAT data with a claim of 5$\sigma$. In this study, we searched for the QPO signature in the entire 15 years (from MJD 54683 to 60187) of $\gamma$-ray light curve and have identified three QPO peaks at frequency 1.24$\times$10$^{-3}$
(806 days), 3.39$\times$10$^{-3}$ (295 days), and 3.73$\times$10$^{-3}$ (268 days) above 3$\sigma$ significance.
We also observed the peak at frequency 0.0022 ($\sim$ 454 days) which is concentrated roughly between MJD 56000 -- 60000 (see the Fig.\ref{fig-2} WWZ plot) but it is much below $3\sigma$ significance in both LSP and WWZ methodology. We believe this could potentially be the 2nd harmonic with a peak at frequency 0.0022 (1/day) ($\sim$454 days) as it has been consistently seen in LSP and WWZ analysis of 15-day and 30-day binned lightcurve but the significance always remains below 3$\sigma$. In addition, the QPO peak observed at 806 days serves as the 3rd harmonics of the fundamental QPO with a period of 268 days. In the literature, several studies have reported the 2nd and 3rd harmonics of fundamental QPO frequency detected below or above a certain 
significant level \citep{wang2014periodic, Bhatta_2020, tripathi2021quasi, banerjee2023detection}. Despite the lower detection significance, these components are considered to play a role in forming a harmonic relationship.

By employing stochastic processes, we analyzed the $\gamma$-ray lightcurve of PKS0521-36, exploring all feasible celerite models and computing the BIC, the values are summarised in appendix Table \ref{tab: model selection}. The optimal celerite model, determined to be model (SHO $\times$ SHO), was then used for fitting the light curve spanning MJD 55358 to 59547. Figure \ref{Fig: fitted lc with celerite} in appendix illustrates the fitting results of lightcurve modeling. The goodness of the fit is evaluated by examining the probability densities of standardized residuals, as well as the autocorrelation function (ACF) of both standardized residuals and the square of standardized residuals. The distribution of standardized residuals is fitted with a normal distribution with the mean close to zero and standard deviation unity, and parameters along with the reduced chi-square ($\chi^2$) are estimated, which is reduced-$\chi^2$ = 2.0025, and presented in Table \ref{tab: Posterior parameters} in appendix. The ACFs of standardized residuals and squared residuals are not fully inside the 95$\%$ confidence intervals of the white noise, which indicates that the model has captured non-linear behaviors in the time series. The stochastic modeling outcomes suggest that the gamma-ray variability is effectively captured by a product of two SHO models. Appendix figure \ref{Fig: posterior parameters} displays the posterior probability densities for the model parameters, while the corresponding best-fitting parameter values are summarized in Table \ref{tab: Posterior parameters} in appendix. Appendix figure \ref{Fig: SHOx2} illustrates the constructed PSD profile, using the parameters obtained through the celerite modeling. The constructed PSD shows two broader peaks at 283 days and 886 days mostly consistent with the observed QPO frequency (295 days, and 806 days) in LSP and WWZ analysis. 

The gamma-ray QPO has been detected in many blazars, thanks to the Fermi-LAT for providing long-term monitoring which makes the QPO detection possible (see e.g. \citealt{2014ApJ...793L...1S, 2015ApJ...810...14A, 2016AJ....151...54S, 2017MNRAS.471.3036P, Bhatta_2020, Das_2023, prince2023quasi}). The QPO signature in blazars is mostly caused by geometrical effects such as jet precession, changing Doppler factor, and blob motion along the curved jet structure.
In general, the jet precession is caused by external effects such as the presence of a supermassive binary black hole system. The QPO time scale in this scenario is  
expected to be the order of years \cite{2015ApJ...810...14A}. However, the radio morphology suggests that in the case of PKS 0521-36, the jet is weakly beamed \citep{2016A&A...586A..70L, 2019A&A...627A.148A}. Another study done by \cite{2017MNRAS.470L.107J} observed a helical motion along the optical jet of PKS 0521-36 and proposed that it could be due to the helical magnetic field structure along the jet or the jet precession. 


Several other possible explanations may emerge to explain our findings in this case study. The observed characteristic periods of $\sim$268 days, $\sim$500 days, and $\sim$806 days are related harmonically and in that way, the observed QPO with period of $\sim$806 days may be the third harmonic of the QPO with a $\sim$268 days period. However, the additional periodicity observed around 295 days remains somewhat enigmatic, lacking a clear association with the other periodicities. The harmonic relationship of the periodicities could be interpreted by periodic variation of orbiting hotspots at the innermost region of the thick disk. In addition, this phenomenon may be attributed to the Kelvin-Helmholtz (K-H) instability, as proposed by \cite{an2013periodic}, which could give rise to global acoustic p-mode oscillations within the inner disk region, subsequently propagating towards the outer regions of the disk. These mechanisms offer promising explanations for the generation of such oscillations.\par

Numerous case studies within the literature have explored the detection of Quasi-Periodic Oscillations across a diverse range of timescales, spanning from mere minutes to hours, days, and extending to months and years. With this wealth of information, we are now equipped to delve into an examination of the correlation between the timescale of QPOs and the black hole masses of AGN. We compiled the gamma-ray QPO time scales from various studies and detailed information is included in Table~\ref{tab:qpo-timescale}. 
In our study, we estimated the correlation between the QPO timescales and their black hole (BH) masses (Figure \ref{fig : QPO_vs_BH_mass} in appendix) 
utilizing the Pearson Correlation coefficient. 
The results of our analysis yielded a correlation coefficient of 0.17 with a p-value of 0.31, suggesting no or a very low correlation. 
The absence of a significant correlation is attributed to the fact that the masses of blazars for which $\gamma$-ray QPOs were observed, fall within a narrow mass range that spans an order of magnitude. To gain a more comprehensive understanding of the underlying emission mechanism responsible for the observed oscillatory features in the light curves of various objects, we propose an expanded exploration of this concept. This involves investigating multi-wavelength detected QPOs and the masses across various astrophysical objects.

\section{CONCLUSION}

We conducted a comprehensive temporal 
search of quasi-periodic oscillations in
$\gamma$-ray emissions from the non-blazar AGN PKS 0521-36, utilizing Fermi-LAT observations spanning an impressive 15-year period, from August 5th, 2008, to August 31st, 2023. Our investigation, employing both LSP and WWZ, revealed the presence of multiple QPOs. These QPOs are identified at a timescale of $\sim$268 days (0.00373 $\text{d}^{-1}$), $\sim$806 days (0.00124 $\text{d}^{-1}$), and $\sim$295 days (0.00339 $\text{d}^{-1}$), with a statistical significance exceeding 3$\sigma$ suggesting the presence of true QPO signature. We also applied the Gaussian Process modeling such as stochastically driven damped harmonic oscillator (SHO) and damped random walk (DRW) to model the light curve. The constructed power spectrum density (PSD) exhibit two broad peak at a timescale of $\sim$283 days and $\sim$886 days mostly consistent with the QPO timescale detected in LSP and WWZ analysis. The observed periods are in harmonic relationship with a 1:3 ratio. 
Furthermore, in LSP and WWZ analysis, we also noted a peak at a frequency of $\sim$0.0022 $\text{d}^{-1}$ (corresponds to a period of $\sim$ 454 days), although its significance couldn't reach the 3$\sigma$ threshold. For the generation of such oscillations, one potential mechanism involves the helical motion of magnetic plasma blobs within the jet or the jet precession. Additionally, due to the observed harmonic relationship for periods, alternative explanations may include the orbital motion of hotspots within the innermost circular thick disk or the presence of K-H instability in the inner disk region.

\section*{Acknowledgements}
We thank the referee for their fruitful comments and suggestions. This research makes use of the publicly available data from Fermi-LAT obtained from the FSSC data server and distributed by NASA Goddard Space Flight Center (GSFC). 
R. Prince is grateful for the support of the Polish Funding Agency National Science Centre, project 2017/26/A/ST9/-00756 (MAESTRO
9) and the European Research Council (ERC) under the European Union’s Horizon 2020 research and innovation program (grant agreement No. [951549]). A. Sharma is grateful to Prof. Sakuntala Chatterjee at S.N. Bose National Centre for Basic Sciences, for providing the necessary support to conduct this research.

\section*{Data Availability}
The work uses publicly available data from \textit{Fermi-}LAT.



\bibliographystyle{mnras}
\bibliography{example} 




\appendix

\section{$\gamma$-ray SED}
\begin{figure}
    \centering
    \includegraphics[width=0.42\textwidth]{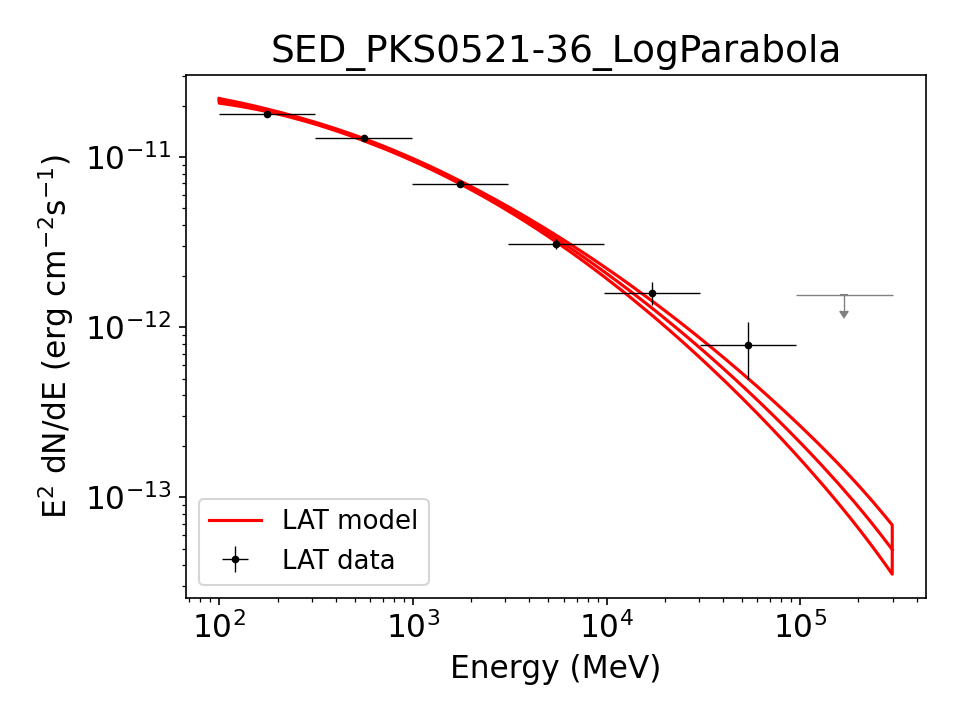}
    \caption{The $\gamma$-ray SED of PKS 0521-36 in the time domain from MJD 54683 – 60187. The best-fitting LogParabola model is plotted as a red solid line along with a 1$\sigma$ contour. The obtained spectral parameters are given in the text.}
    \label{Fig: GAMMA-RAY SED}
\end{figure}

The $\gamma$-ray SED is fitted with a log-parabola spectal model and shown in Fig.~\ref{Fig: GAMMA-RAY SED}.

\section{Gaussian Modeling}
\begin{figure}
    \centering
    \includegraphics[width=0.52\textwidth]{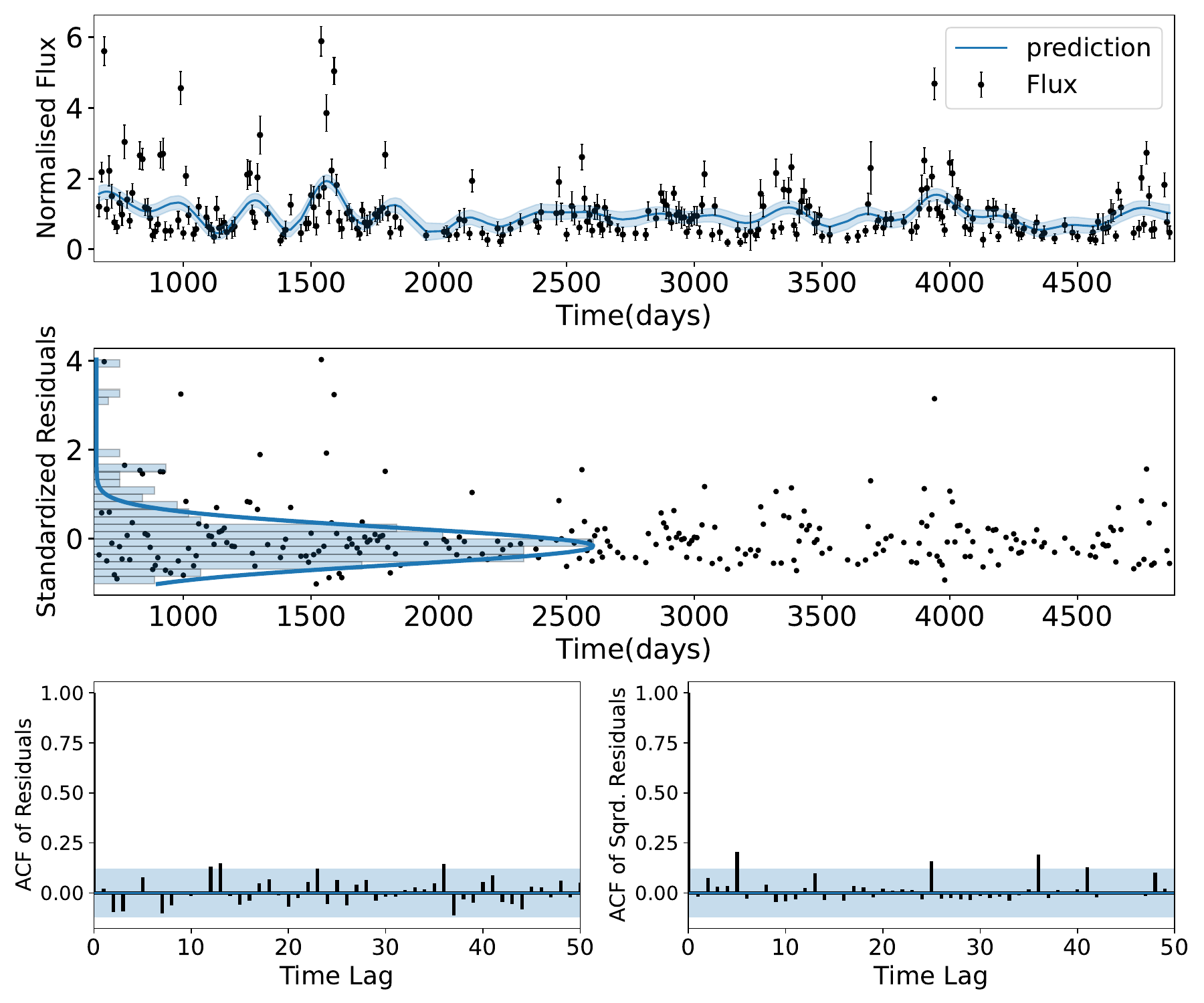}
    \caption{The celerite modeling was applied to a 10-day binned $\gamma$-ray light curve of PKS 0521-36 spanning from MJD 55358 to 59547, employing the SHO×SHO model. The top panel illustrates the flux points and associated uncertainties in black, while the best-fit profile resulting from the celerite modeling, along with its 1$\sigma$ confidence interval, is depicted in blue. In the middle panel, black dots represent the standardized residuals of each bin, and a histogram of scaled standardized residuals is overlaid in blue, accompanied by a solid blue line denoting the expected scaled normal distribution. The bottom panel showcases the Auto-Correlation Functions (ACFs) of standardized residuals and squared standardized residuals, along with 95$\%$ confidence intervals of the white noise.   }
    \label{Fig: fitted lc with celerite}
\end{figure}

\begin{figure}
    \centering
    \includegraphics[width=0.52\textwidth]{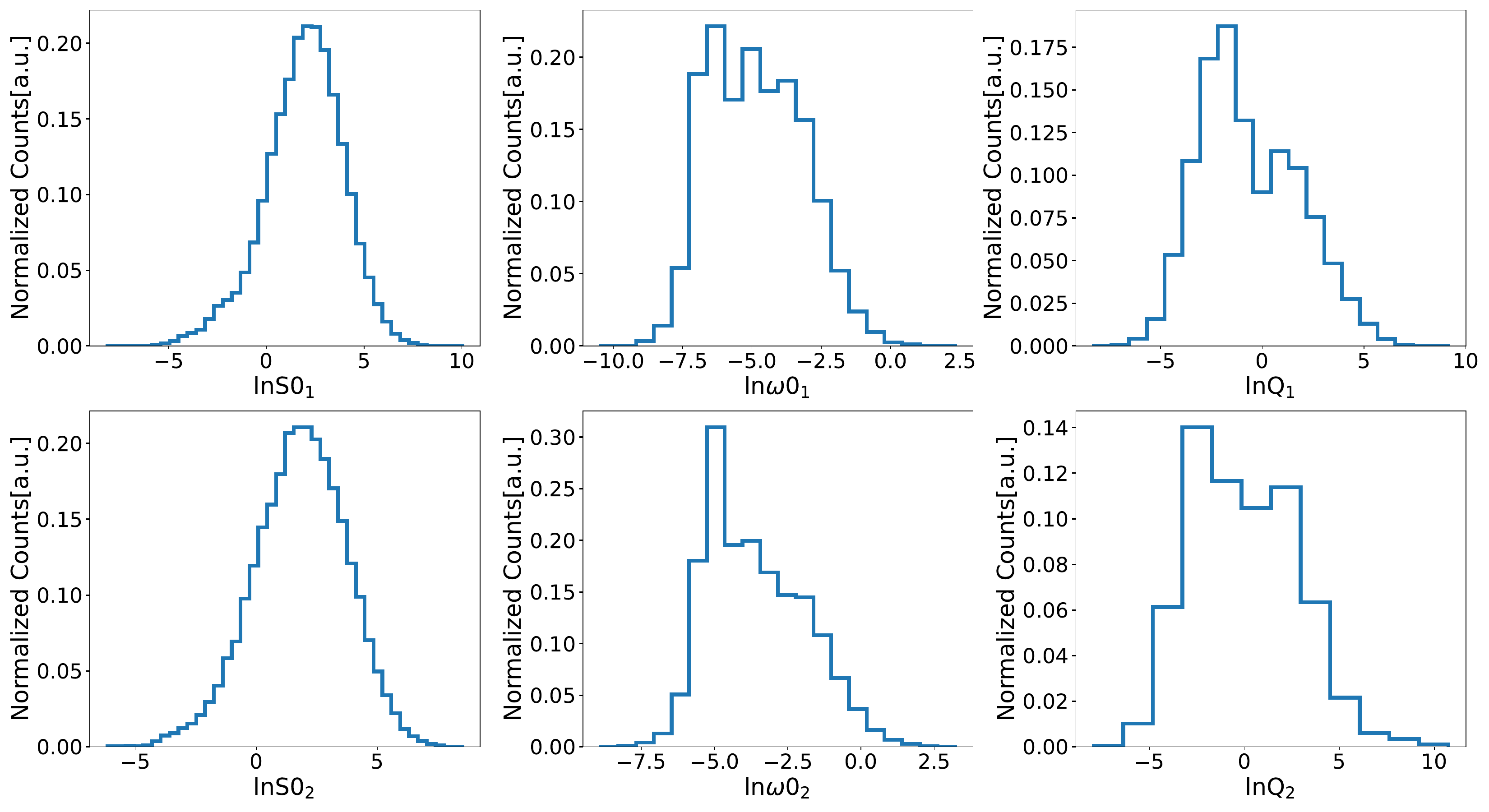}
    \caption{The posterior probability densities of parameters obtained from celerite modeling with SHO$\times$SHO model.}
    \label{Fig: posterior parameters}
\end{figure}

\begin{figure}
    \centering
    \includegraphics[width=0.45\textwidth]{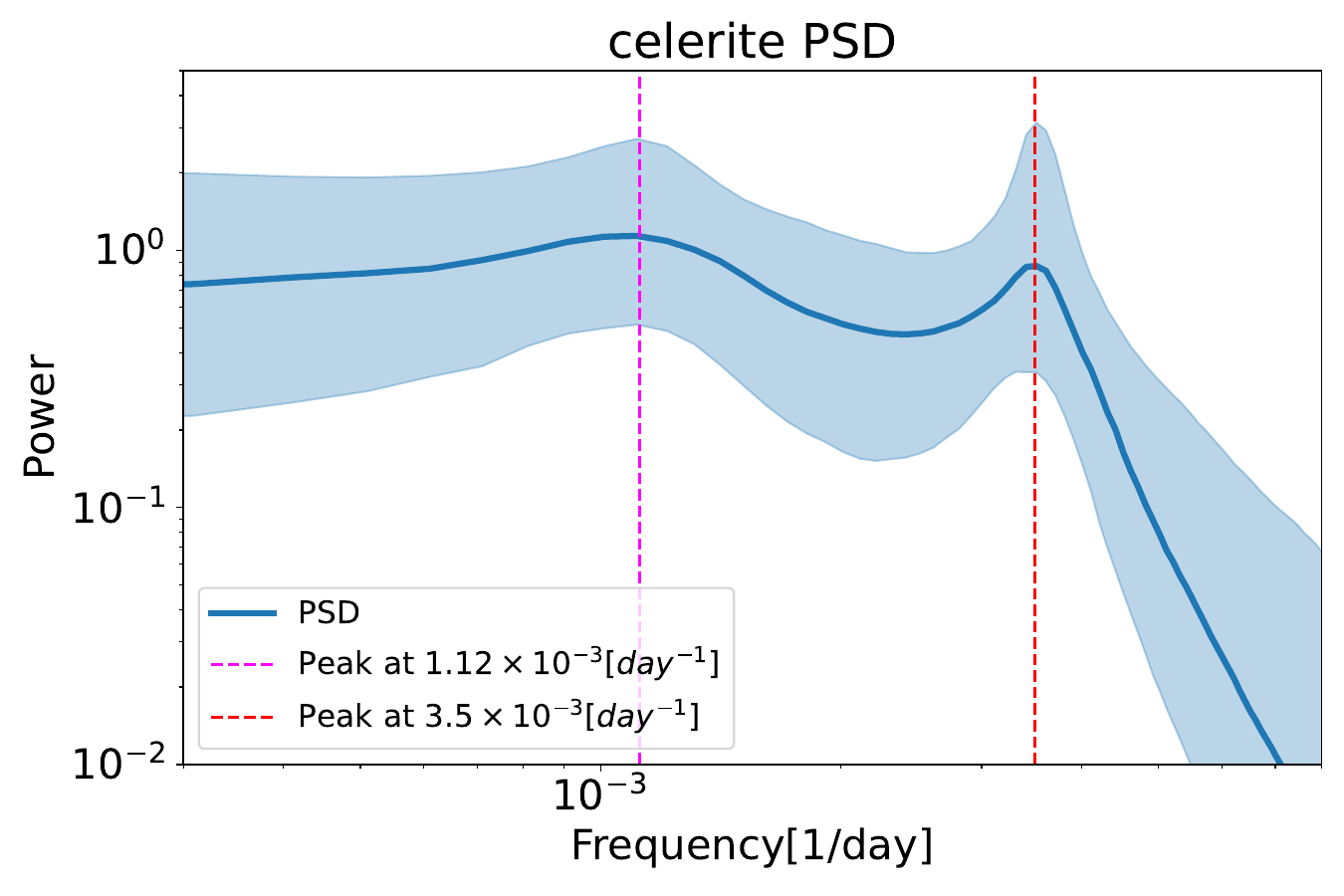}
    \caption{The PSD constructed from the celerite modeling results with SHO$\times$SHO model for the 10-day binned $\gamma$-ray light curve of PKS 0521-36 during MJD 55358 to 59547. The discovered QPOs features at periods of $\sim$283 days and $\sim$886 days with the 68$\%$ confidence band.}
    \label{Fig: SHOx2}
\end{figure}

Various combinations of SHO and DRW models are used to optimize for the best variability model to explain the gamma-ray light curve. The corresponding parameters are shown in Table~\ref{tab: model selection} \& \ref{tab: Posterior parameters}. The selected best model is SHO$\times$SHO and the corresponding plots are shown in Fig~\ref{Fig: fitted lc with celerite} \& \ref{Fig: posterior parameters}. Using the best fit parameters the PSD is constructed and shown in Fig~\ref{Fig: SHOx2}.



\begin{table}
\centering

\setlength{\extrarowheight}{4pt}
\begin{tabular}{| c |ccc|}

\hline
\multirow{2}{*}{\large{Source}} & \multirow{2}{*}{Model} & \multirow{2}{*}{Log likelihood} & \multirow{2}{*}{BIC}  \\
(1) & (2) & (3) & (4) \\
\hline

\multirow{12}{*}{PKS 0521-36} & SHO & -297.09 & 627.55 \\
& SHO$\times$2 & -295.96 & 625.28 \\
& SHO$\times$3 & -294.37 & 638.80 \\
& SHO$\times$4 & -289.57 & 645.88 \\
[+4pt]
\cline{2-4}
& SHO & -297.092 & 627.55 \\
& SHO + 2 & -297.83 & 629.02 \\
& SHO + 3 & -294.20 & 638.45 \\
& SHO + 4 & -292.91 & 652.56 \\
[+4pt]
\cline{2-4}
& DRW + SHO & -300.97 & 629.75 \\
& DRW + SHO$\times$2 & -295.30 & 635.10 \\
& DRW + SHO$\times$3 & -295.70 & 652.57 \\
& DRW + SHO$\times$4 & -289.59 & 657.03\\
[+4pt]
\hline
\end{tabular}

\caption{\label{tab: model selection} The optimal model selection for the stochastic process modeling of $\gamma$-ray lightcurve. (1) source name, (2) Model, (3) loglikelihood value, and (4) the value of Bayesian Information Criterion (BIC) for a model. These suggest the SHO$\times$2 is the best model to explain the variability in the $\gamma$-ray light curve.}
\end{table}


\begin{table}
\centering
\setlength{\extrarowheight}{4pt}
\begin{tabular}{ccccc}
\hline
\hline
\multirow{2}{*}{\large{Source}} & \multirow{2}{*}{Model} & \multirow{2}{*}{ln $S_0$} & \multirow{2}{*}{ln Q} & \multirow{2}{*}{ln $\omega_0$}  \\
(1) & (2) & (3) & (4) & (5) \\
\hline

\multirow{2}{*}{PKS 0521-36} & \multirow{2}{*}{SHO$\times$SHO} & $2.04_{-2.01}^{+1.79}$ & $-1.12_{-1.92}^{+3.16}$ & $-4.95_{-0.95}^{+1.30}$ \\
& & $1.88_{-1.93}^{+1.83}$ & $-0.27_{-2.54}^{+3.11}$ & $-3.81_{-0.76}^{+0.82}$ \\
[+4pt]
\hline
\end{tabular}
\caption{\label{tab: Posterior parameters} The best-fitting parameters of the SHO$\times$2 model are shown here. (1) source name, (2) Model, (3)-(5) posterior parameters of the SHO$\times$2 model.}
\end{table}


\section{QPO time scale versus BH mass}
\begin{figure}
    \centering
    \includegraphics[width=0.45\textwidth]{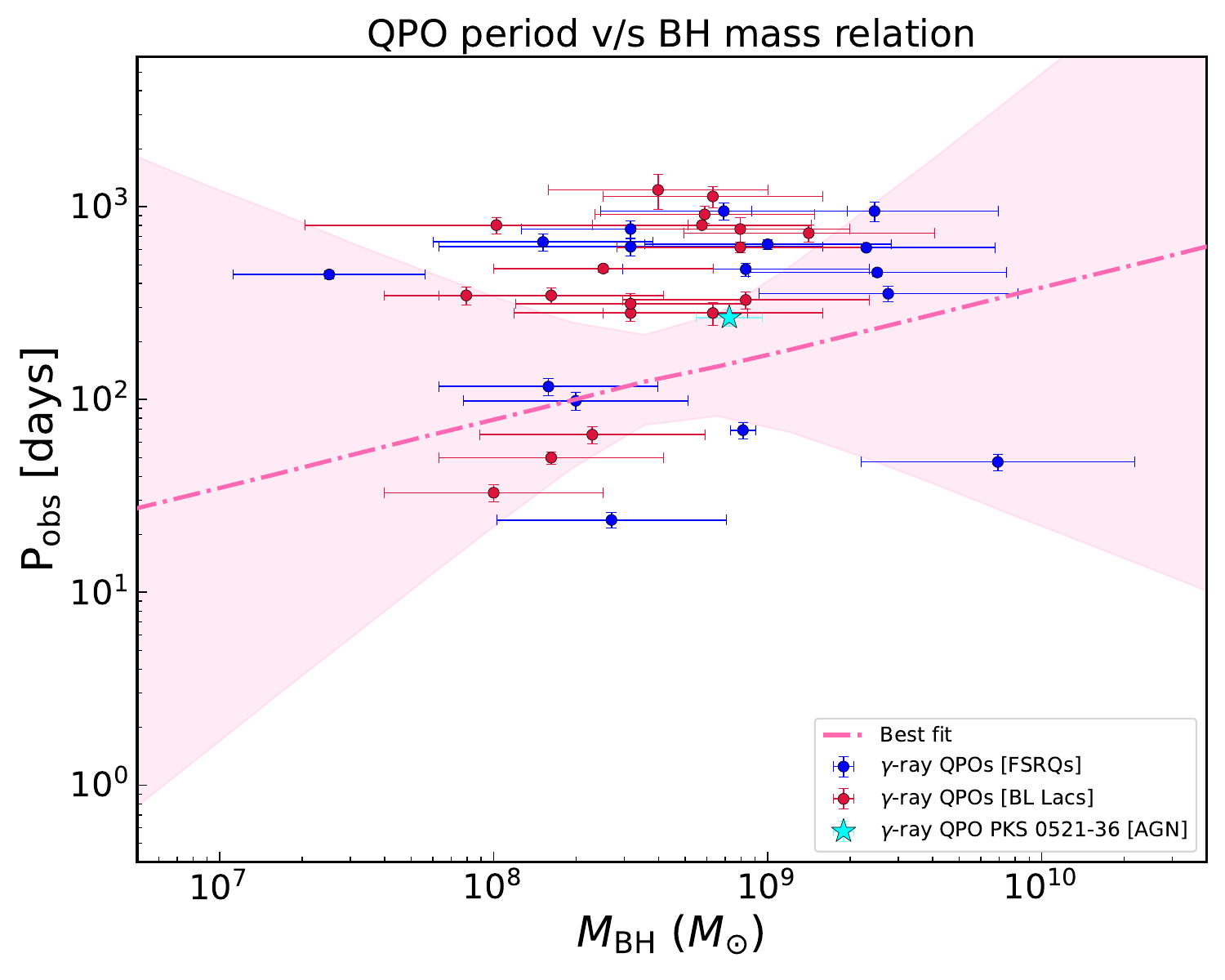}
    \caption{The observed QPO timescales and their relationship with black hole (BH) mass, Table \ref{tab: qpo-timescale}. All the blue points represent the FSRQs, while BL Lacs are illustrated in red. The source PKS 0521-36 is mentioned by a star symbol in cyan color. A fit with its 68$\%$ confidence region is depicted in pink. }
    \label{fig : QPO_vs_BH_mass}
\end{figure}

The collected sample of gamma-ray detected QPO and their time scales are plotted along with the BH mass of the objects are shown in Fig~\ref{fig : QPO_vs_BH_mass}. No significant correlation is observed.

\begin{table*} \footnotesize
\centering
\setlength{\extrarowheight}{6pt}
\begin{tabular}{ccccccccc}
\hline
\hline
\multirow{2}{*}{Source Name} & \multirow{2}{*}{4FGL name} & \multirow{2}{*}{R.A.} & \multirow{2}{*}{Dec.} & \multirow{2}{*}{Source Type} & \multirow{2}{*}{\textit{z}} & \multirow{2}{*}{QPO timescale (yr)} & \multirow{2}{*}{log($M/M_{\odot}$)} & \multirow{2}{*}{Reference}\\
(1) & (2) & (3) & (4) & (5) & (6) & (7) & (8) & (9) \\
\hline
GB6 J0043+3426 & 4FGL J0043.8+3425 & 10.9717 & 34.4316 & FSRQ & 0.966 & 1.8 & 8.18 & (1) \\
TXS 0059+581 & 4FGL J0102.8+5824 & 15.701 & 58.4092 & FSRQ & 0.644 & 2.1 & 8.5 & (1) \\
PKS 0208$-$512 & 4FGL J0210.7$-$5101 & 32.6946 & $-$51.0218 & FSRQ & 1.003 & 2.6 & 8.84 & (1) \\
PKS 0250$-$225 & 4FGL J0252.8$-$2219 & 43.2007 & $-$22.3203 & FSRQ & 1.41 & 1.25 & 9.4 & (1) \\
PKS 0454$-$234 & 4FGL J0457.0$-$2324 & 74.2608 & $-$23.4149 & FSRQ & 1.003 & 2.6 & 9.39 & (1) \\
4C $-$02.19 & 4FGL J0501.2$-$0158  & 75.3023 & $-$1.9749 & FSRQ & 2.286 & 1.7 & 8.5 & (1) \\
PKS 2052$-$477 & 4FGL J2056.2$-$4714 & 314.0715 & $-$47.23 & FSRQ & 1.49 & 1.75 & 9 & (5) \\
PKS 2255$-$282 & 4FGL J2258.1$-$2759 & 344.5288 & $-$27.9843 & FSRQ & 0.926 & 1.3 & 8.92 & (1) \\
PKS 0346-27 & 4FGL J0348.5-2749 & 57.1589 & -27.8204 & FSRQ & 0.991 & 0.27 & 8.3 & (15) \\
PKS 0601-70 & 4FGL J0601.1-7035 &  90.2969 & -70.6024  & FSRQ & 2.409 & 1.22 & 7.4 & (9) \\
3C 454.3 & 4FGL J2253.9+1609 & 343.491 & 16.1482 & FSRQ & 0.85 &0.13 & 9.84 & (10) \\
B2 1520+31 & 4FGL J1522.1+3144 & 230.542 & 31.7373 &FSRQ & 1.48 & 0.19 & 8.91 & (10) \\
PKS 1510-089 & 4FGL J1512.8-0906 & 228.211 & -9.09995 & FSRQ & 0.35 & 0.32 & 8.20 & (10) \\
PKS 1424-418 & 4FGL J1427.9-4206 & 216.985 & -42.1054 & FSRQ & 1.52 & 0.97 & 9.44 & (10) \\
3C 279 & 4FGL J1256.1-0547 & 194.047 & -5.78931 & FSRQ & 0.53 & 0.065 & 8.43 & (10) \\
B2 1633+38 & 4FGL J1635.2+3808 & 248.815 & 38.1346 & FSRQ & 1.81 & 1.68 & 9.36 & (11) \\
PKS 0301$-$243 & 4FGL J0303.4$-$2407 & 45.8625 & $-$24.1225 & BLL & 0.266 & 2.1 & 8.9 & (2) \\
PKS 0426$-$380 & 4FGL J0428.6$-$3756 & 67.173 & $-$37.9403 & BLL & 1.11 & 3.35 & 8.6 & (3) \\
PKS 0447$-$439 & 4FGL J0449.4$-$4350 & 72.3582 & $-$43.835 & BLL & 0.205 & 2.5 & 8.77 & (1) \\
TXS 0518$+$211 & 4FGL J0521.7$-$2112 & 80.4445 & 21.2131 & BLL & 0.108 & 3.1 & 8.8 & (1) \\
PKS 0537$-$441 & 4FGL J0538.8$-$4405 & 84.7089 & $-$44.0862 & BLL & 0.896 & 0.77 & 8.8 & (4) \\
S5 0716+714 & 4FGL J0721.9+7120 & 110.4882 & 71.3405 & BLL & 0.31 & 0.95 & 7.9 & (5) \\
S4 0814+42 & 4FGL J0818.2+4222 & 124.56174 & 42.38367 & BLL & 0.53 & 2.2 & 8.01 & (1)\\
PG 1553+113 & 4FGL J1555.7+1111 & 238.9313 & 11.1884 & BLL & 0.49 & 2.2 & 8.76 & (6) \\ 
PKS 2155$-$304 & 4FGL J2158.8$-$3013 & 329.7141 & $-$30.2251 & BLL & 0.116 & 1.69 & 8.9 & (7,8) \\
BL Lacertae & 4FGL J2202.7+4216 & 330.6946 & 42.2821 & BLL & 0.069 & 0.137 , 0.95 & 8.21 & (5,8,14) \\
B2 2234+28 A & 4FGL J2236.3+2828 & 339.094 & 28.4826 & BLL & 0.79 & 1.31 & 8.4 & (9) \\
PKS 2247-131 & 4FGL J2250.0-1250 & 342.498 & -12.8547 & BLL & 0.22 & 0.09 & 8.00 & (10) \\
Mrk 501 & 4FGL J1653.8+3945 & 253.468 & 39.7602 & BLL & 0.034 & 0.9 & 8.92 & (10, 11) \\
Mrk 421 & 4FGL J1104.4+3812 & 166.114 & 38.2088 & BLL & 0.03 & 0.77 & 8.5 & (10, 11) \\
PG 1246+586 & 4FGL J1248.3+5820 & 192.078 & 58.3413 & BLL & 0.84 & 2.0 & 9.15 & (1) \\
OJ 287 & 4FGL J0854.8+2006 & 133.704 & 20.1085 & BLL & 0.306 & 0.86 & 8.5 & (12) \\
S4 0954+65 & 4FGL J0958.7+6534 & 149.697 & 65.5652 & BLL & 0.36 & 0.18 & 8.36 & (13) \\
\textbf{PKS 0521-36} & \textbf{4FGL J0522.9-3628} & \textbf{80.7416} & \textbf{-36.4586} & \textbf{AGN} & \textbf{0.055} & \textbf{0.73} & \textbf{8.86} & this work \\
\hline
\end{tabular}
\caption{\label{tab: qpo-timescale} The information of 33 blazars and 1 AGN (source of interest) is given here. (1) source name, (2) 4FGL source name, (3)-(4) Coordinate of source, (5) Source Type, (6) Redshift, (7) QPO timescale (in year), (8) Black Hole mass (in solar mass), and (9) all the information has been assembled from the listed references : (1) \citealt{penil2020systematic}, (2) \citealt{zhang2017gamma}, (3) \citealt{zhang2017possible}, (4) \citealt{sandrinelli2016gamma}, (5) \citealt{prokhorov2017search}, (6) \citealt{ackermann2015multiwavelength}, (7) \citealt{zhang2017revisiting}, (8) \citealt{sandrinelli2018quasi}, (9) \citealt{zhang2020searching}, (10) \citealt{ren2023quasi}, (11) \citealt{tarnopolski2020comprehensive}, (12) \citealt{kushwaha2020possible}, (13) \citealt{gong2023two}, (14) \citealt{banerjee2023detection}, (15) \citealt{prince2023quasi}  }
\end{table*}

\label{lastpage}
\end{document}